\documentclass[a4paper,11pt]{article}
\pdfoutput=1 

\usepackage{jcappub} 

\usepackage{color}
\usepackage{amsmath}

\title{\boldmath Parameter estimation with Sandage-Loeb test}

\author[a]{Jia-Jia Geng,}
\author[a]{Jing-Fei Zhang,}
\author[a,b,1]{Xin Zhang\note{Corresponding author.}}

\affiliation[a]{Department of Physics, College of Sciences, Northeastern University, \\Shenyang
110004, China}
\affiliation[b]{Center for High Energy Physics, Peking University, \\Beijing 100080, China}

\emailAdd{gengjiajia163@163.com}
\emailAdd{jfzhang@mail.neu.edu.cn}
\emailAdd{zhangxin@mail.neu.edu.cn}

\abstract{
The Sandage-Loeb (SL) test directly measures the expansion rate of the universe in the redshift range of $2\lesssim z\lesssim 5$ by detecting redshift drift in
the spectra of Lyman-$\alpha$ forest of distant quasars.
We discuss the impact of the future SL test data on parameter estimation for the $\Lambda$CDM, the $w$CDM, and the $w_0w_a$CDM models.
To avoid the potential inconsistency with other observational data, we take the best-fitting dark energy model constrained by the current observations as the fiducial model
to produce 30 mock SL test data.
The SL test data provide an important supplement to the other dark energy probes, since they are extremely helpful in breaking the existing parameter degeneracies.
We show that the strong degeneracy between $\Omega_m$ and $H_0$ in all the three dark energy models is well broken by the SL test.
Compared to the current combined data of type Ia supernovae, baryon acoustic oscillation, cosmic microwave background, and Hubble constant, the 30-yr observation of
SL test could improve the constraints on $\Omega_m$ and $H_0$ by more than 60\% for all the three models.
But the SL test can only moderately improve the constraint on the equation of state of dark energy.
We show that a 30-yr observation of SL test could help improve the constraint on constant $w$ by about 25\%, and improve the constraints on $w_0$ and $w_a$ by
about 20\% and 15\%, respectively.
We also quantify the constraining power of the SL test in the future high-precision joint geometric constraints on dark energy.
The mock future supernova and baryon acoustic oscillation data are simulated based on the space-based project JDEM.
We find that the 30-yr observation of
SL test would help improve the measurement precision of $\Omega_m$, $H_0$, and $w_a$ by more than 70\%, 20\%, and 60\%, respectively, for the $w_0w_a$CDM model.
}

\begin{document}
\maketitle
\flushbottom
\section{Introduction}

Sandage-Loeb (SL) test is a unique method to directly measure the expansion history of the universe in the ``redshift desert'' of $2\lesssim z\lesssim 5$. It was firstly proposed by Sandage~\cite{sandage} to directly measure the variation of redshift of distant sources. Then Loeb~\cite{loeb} pointed out the possibility of detecting redshift drift in the spectra of Lyman-$\alpha$ forest of distant quasars (QSO) in decades. The 39-meter European Extremely Large Telescope (E-ELT) equipped with a high-resolution spectrograph called CODEX (COsmic Dynamics EXperiment) is in built to achieve this goal. The SL test is of great significance for cosmology because it is a direct geometric measurement of the expansion history of the universe and covers the high redshift range of $2\lesssim z\lesssim 5$, which is almost unaccessible with existing probes.

The effect of the SL test on parameter estimation has been studied by enormous works~\cite{sl1,sl2,sl3,sl4,sl5,sl6,sl7,Darling,Zhang21}, however, many works incorrectly assumed 240 or 150 quasars to be observed. In fact, according to a Monte Carlo simulation analyzed in depth, using a telescope with a spectrograph like CODEX, only about 30 quasars are bright enough and/or lying at a high enough redshift for the actual observation~\cite{Liske}. Moreover, as far as we know, in almost all the existing papers,  the best-fit $\Lambda$CDM model to current observational data is usually chosen as the fiducial model in simulating the mock future SL test data. In such a way, when these simulated data are combined with other actual data to constrain some dynamical dark energy models (or modified gravity models), tension between the simulated SL data and other actual data may occur, leading to an inappropriate joint constraint. Thus, such a method may not give convincing conclusion on the potential impact of the future SL test data on parameter estimation.

In our recent work~\cite{msl1}, we suggested that to avoid the potential inconsistency in data the best-fitting model (in study) to current actual data is taken to be the fiducial model in
producing the simulated SL test data, and 30 mock data are then produced with this procedure. In such a way, the simulated mock data are well consistent with the current actual data no matter what dark energy
models are considered. The conclusion of the impact of SL test on future parameter estimation is thus rather convincing. In Ref.~\cite{msl1}, as a typical example, we only focused on the
dark energy model with constant $w$ (i.e., the $w$CDM model).
It was shown that compared to the current combined data of type Ia supernovae (SN), baryon acoustic oscillation (BAO), cosmic microwave background (CMB), and Hubble constant, the 30-yr observation of SL
test could improve the constraint on $\Omega_m$ by about 80\% and the constraint on $w$ by about 25\%.
Furthermore, if the interaction between dark energy and dark matter is considered, the SL test 30-yr data could also improve the constraint on the coupling $\gamma$ by about 30\% and 10\%
for the $Q=\gamma H\rho_{\rm c}$ and $Q=\gamma H\rho_{\rm de}$ models, respectively, as shown in Ref.~\cite{msl1}.

In this paper, we will further extend the discussions in Ref.~\cite{msl1} and investigate the parameter estimation with the SL test in depth.
We will consider the case of time-evolving dark energy model, and show how the SL test impacts on the constraints on the equation of state of
such a dark energy. As usual, we adopt the most commonly used parametrization $w(z)=w_0 + w_az/(1+z)$, and call the corresponding model the $w_0w_a$CDM model.
A comprehensive comparison among the $\Lambda$CDM, the $w$CDM, and the $w_0w_a$CDM models with the SL test will be performed.
Another important issue is about the determination of the Hubble constant by using the future SL test data.
It is well known that in the current data there is a strong degeneracy between $\Omega_m$ and $H_0$ (they are in an anti-correlation).
Breaking this degeneracy is extremely important for cosmology.
In this work, we will show that the SL test is very helpful in breaking the degeneracy between $\Omega_m$ and $H_0$, and thus
is very helpful in determining the value of the Hubble constant.
Furthermore, we will also discuss what accuracy would be achieved when using the SL test to directly measure the high-redshift $H(z)$ values.

In fact, a more meaningful question is to ask how the SL test would impact on the dark energy constraints in the future geometric measurements.
We will also address this issue in the present work. As a concrete example, we simulate the future SN and BAO data based on the long-term space-based project JDEM.
We wish to quantify the constraining power of the SL test in the future high-precision joint geometric constraints on dark energy.


\section{Methodology}

First, we briefly describe the current observational data used in the analysis. Actually, the current data used in this work are the same to those in Ref.~\cite{msl1},
in order to make a direct comparison. The most typical geometric measurements are chosen, i.e., the observations of SN, BAO, CMB, and $H_0$.
The combination of SN, BAO, CMB, and $H_0$ is, actually, the most commonly used data combination in parameter estimation studies of dark energy models.
For the SN data, the SNLS compilation~\cite{snls3} with a sample of 472 SNe is used in this work.
For the BAO data, we consider the $r_s/D_V(z)$ measurements from 6dFGS ($z=0.1$), SDSS-DR7 ($z=0.35$), SDSS-DR9 ($z=0.57$), and
WiggleZ ($z=0.44$, 0.60, and 0.73) surveys, where the three data from the WiggleZ survey are correlated (for the data and their inverse covariance matrix,
see, e.g., Ref.~\cite{wmap9}).
For the CMB data, we adopt the Planck distance posterior given by Ref.~\cite{WW}.
It should be noted that dark energy only affects the CMB through the comoving angular diameter distance to the decoupling epoch (and the late-time ISW effect),
and so the distance information given by the CMB distance posterior is sufficient for the joint geometric constraint on dark energy.
We also use the direct measurement result of the Hubble constant in the light of the cosmic distance ladder from the HST, $H_0=73.8\pm 2.4$ km s$^{-1}$ Mpc$^{-1}$~\citep{Riess2011}.

Our procedure is as follows. Dark energy models are first constrained by using the current joint SN+BAO+CMB+$H_0$ data, and then
the best-fit dark energy models are chosen to be the fiducial models in producing the simulated mock SL test data.
The obtained simulated SL test data are thus well consistent with the current SN+BAO+CMB+$H_0$ data.
Therefore, it is rather appropriate to combine the mock SL test data with the current SN, BAO, CMB, and $H_0$ data
for further constraining dark energy models.

Next, we briefly review how to simulate the SL test data.
This method is just to directly measure the redshift variation of quasar Lyman-$\alpha$ absorption lines.
The redshift variation is defined as a spectroscopic velocity shift \cite{loeb},
\begin{equation}\label{eq6}
\ \Delta v \equiv \frac{\Delta z}{1+z}=H_0\Delta t_o\bigg[1-\frac{E(z)}{1+z}\bigg],
\end{equation}
where $\Delta t_o$ is the time interval of observation, and $E(z)=H(z)/H_0$ is given by specific dark energy models. In a flat universe, we have
\begin{equation}\label{Ez}
E(z)=\sqrt{\Omega_r(1+z)^4+\Omega_m(1+z)^3+(1-\Omega_r-\Omega_m){X}(z)},
\end{equation}
where $\Omega_r$ and $\Omega_m$ are the present-day density parameters of radiation and matter, respectively, and ${X}(z)\equiv\rho_{\rm de}(z)/\rho_{\rm de}(0)=\exp[3\int_0^z {1+w(z')\over 1+z'}dz']$.

According to the Monte Carlo simulations, the uncertainty of $\Delta v$ measurements expected by CODEX can be expressed as~\cite{Liske}
\begin{equation}\label{eq7}
\sigma_{\Delta v}=1.35
\bigg(\frac{S/N}{2370}\bigg)^{-1}\bigg(\frac{N_{\mathrm{QSO}}}{30}\bigg)^{-1/2}\bigg(
\frac{1+z_{\mathrm{QSO}}}{5}\bigg)^{f}~\mathrm{cm}~\mathrm{s}^{-1},
\end{equation}
where $S/N$ is the signal-to-noise ratio defined per 0.0125 ${\AA}$ pixel, $N_{\mathrm{QSO}}$ is the
number of observed quasars, $z_{\mathrm{QSO}}$ represents their
redshift, and the last exponent $f=-1.7$ for $2<z<4$ and $f=-0.9$ for $z>4$. We choose $N_{\mathrm{QSO}}=30$ mock SL data uniformly distributed among
six redshift bins of $z_{\rm QSO}\in [2,~5]$.

To simulate the SL test data, we first constrain the dark energy models by using the current SN+BAO+CMB+$H_0$ data.
We perform an MCMC likelihood analysis~\cite{cosmomc} to obtain ${\cal O}(10^6)$ samples for each model.
The obtained best-fit parameters are substituted into Eq.~(\ref{eq6}) to get the central values of the SL test data,
and we typically take $\Delta t_o=20$ and 30 yr, in our analysis.
The error bars are directly computed from Eq.~(\ref{eq7}) with $S/N=3000$.

\section{Results and discussion}

\begin{table}
\caption{Fit results for the $\Lambda$CDM, $w$CDM, and $w_0w_a$CDM models using the
current CMB+BAO+SN+$H_0$ data.}
\label{table1}
\renewcommand{\arraystretch}{1.5}
\centering
\begin{tabular}{cccccccccc}
\\
\hline\hline
Parameter  & $\Lambda$CDM & $w$CDM & $w_0w_a$CDM\\ \hline

$\Omega_bh^2$      & $0.02237^{+0.00025}_{-0.00024}$
                   & $0.02218^{+0.00025}_{-0.00029}$
                   & $0.02221^{+0.00028}_{-0.00028}$
                   \\

$\Omega_ch^2$      & $ 0.1174^{+0.0014}_{-0.0016}$
                   & $0.1201^{+0.0020}_{-0.0022}$
                   & $0.1194^{+0.0026}_{-0.0027}$
                   \\
$w_0$              & $-1~({\rm fixed})$
                   & $-1.103^{+0.058}_{-0.058}$
                   & $-1.158^{+0.133}_{-0.115}$
                   \\

$w_a$              & $0~({\rm fixed})$
                   & $0~({\rm fixed})$
                   & $0.271^{+0.425}_{-0.638}$
                   \\

$\Omega_{m}$       & $0.2953^{+0.0084}_{-0.0092}$
                   & $0.2844^{+0.0104}_{-0.0093}$
                   & $0.2847^{+0.0089}_{-0.0116}$
                   \\

$H_0$              & $68.81^{+0.75}_{-0.66}$
                   & $70.74^{+1.26}_{-1.30}$
                   & $70.52^{+1.40}_{-1.02}$
                   \\

\hline
\end{tabular}
\end{table}

In this work, we make a comparison for the $\Lambda$CDM, the $w$CDM, and the $w_0w_a$CDM models in the cosmological parameter constraints
with the SL test. First, we constrain the three dark energy models by using the current CMB+BAO+SN+$H_0$ data combination.
Detailed fit results are given in Table~\ref{table1}.
Indeed, as indicated in Ref.~\cite{hdeLi}, when a dynamical dark energy model is considered, the value of $H_0$ will become larger, relieving the tension between
Planck data and $H_0$ direct measurement.
Using the best-fit parameters given in Table~\ref{table1}, the SL test data for constraining each model can be simulated and will be used in the analysis.

\begin{figure}
\begin{center}
\includegraphics[width=13cm]{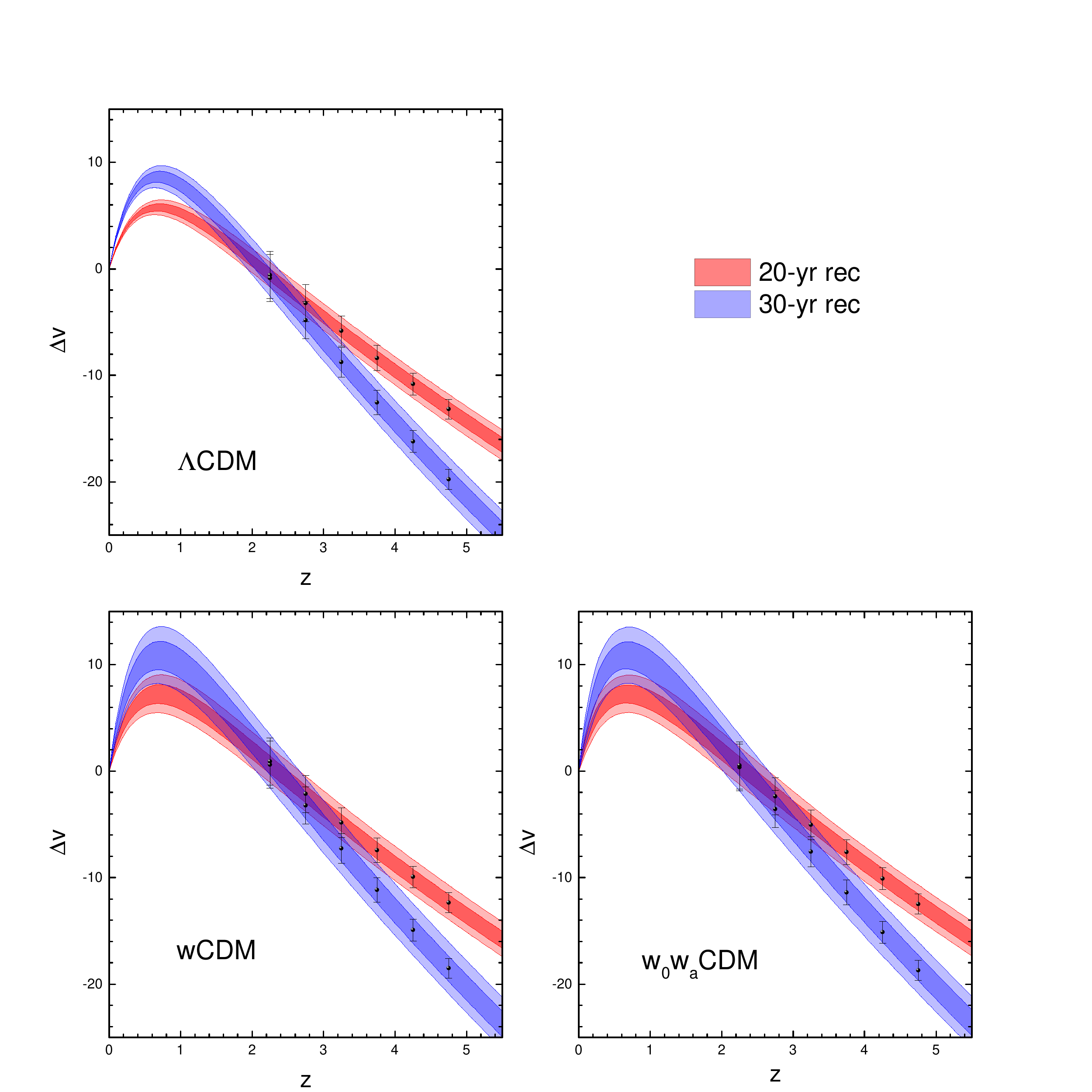}
\end{center}
\caption{Reconstructed redshift drift using the current
SN+BAO+CMB+$H_0$ constraint results for 20-yr and 30-yr observations of SL test.  In this plot, 1$\sigma$ and 2$\sigma$ uncertainties are shown. Error bars (1$\sigma$) of the simulated SL test mock data are also shown for a direct comparison.}
\label{fig1}
\end{figure}

To directly compare the accuracies of the current actual data with the future SL test data, we reconstruct the velocity shifts for the three dark energy models by using the
fit results given in Table~\ref{table1}, plotted in Fig.~\ref{fig1}, as colored bands.
These bands are obtained using the Monte Carlo method. Based on the parameter spaces constrained from the current data combination, the boundaries of $\Delta v$ could be  determined by using Eq.~(\ref{eq6}).
Red and blue bands are for the 20-yr and 30-yr velocity-shift reconstructions, respectively.
We also plot the error bars in the SL test, given by Eq.~(\ref{eq7}), on the corresponding bands, in order to make a direct comparison with the reconstructed results from the current data.
The case of the $w$CDM model has been discussed in Ref.~\cite{msl1}. Now, one can direct compare the three dark energy models.
The conclusion is the same. The 20-yr SL observation would significantly improve the
accuracy; a 30-yr SL observation would be closer to the current combined observations in accuracy, implying that the SL test as a high-redshift supplement to other geometric measurements
will play a crucial role in the parameter estimation in the forthcoming decades.\footnote{To be more quantitative, we
take the velocity shift $\Delta v$ at $z=4.5$ as an example: the ratio of $1\sigma$ uncertainty from SL-20yr data to that from current data, $\sigma_{\rm SL}/\sigma_{\rm current}$, is $1.45$, $1.24$, and $1.24$ for the $\Lambda$CDM, the $w$CDM, and the $w_0w_a$CDM model, respectively, while the ratio of $1\sigma$ uncertainty from SL-30yr data to that from current data, $\sigma_{\rm SL}/\sigma_{\rm current}$, is $0.97$, $0.82$, and $0.83$ for the $\Lambda$CDM, the $w$CDM, and the $w_0w_a$CDM model, respectively.}

\begin{figure}
\begin{center}
\includegraphics[width=15cm]{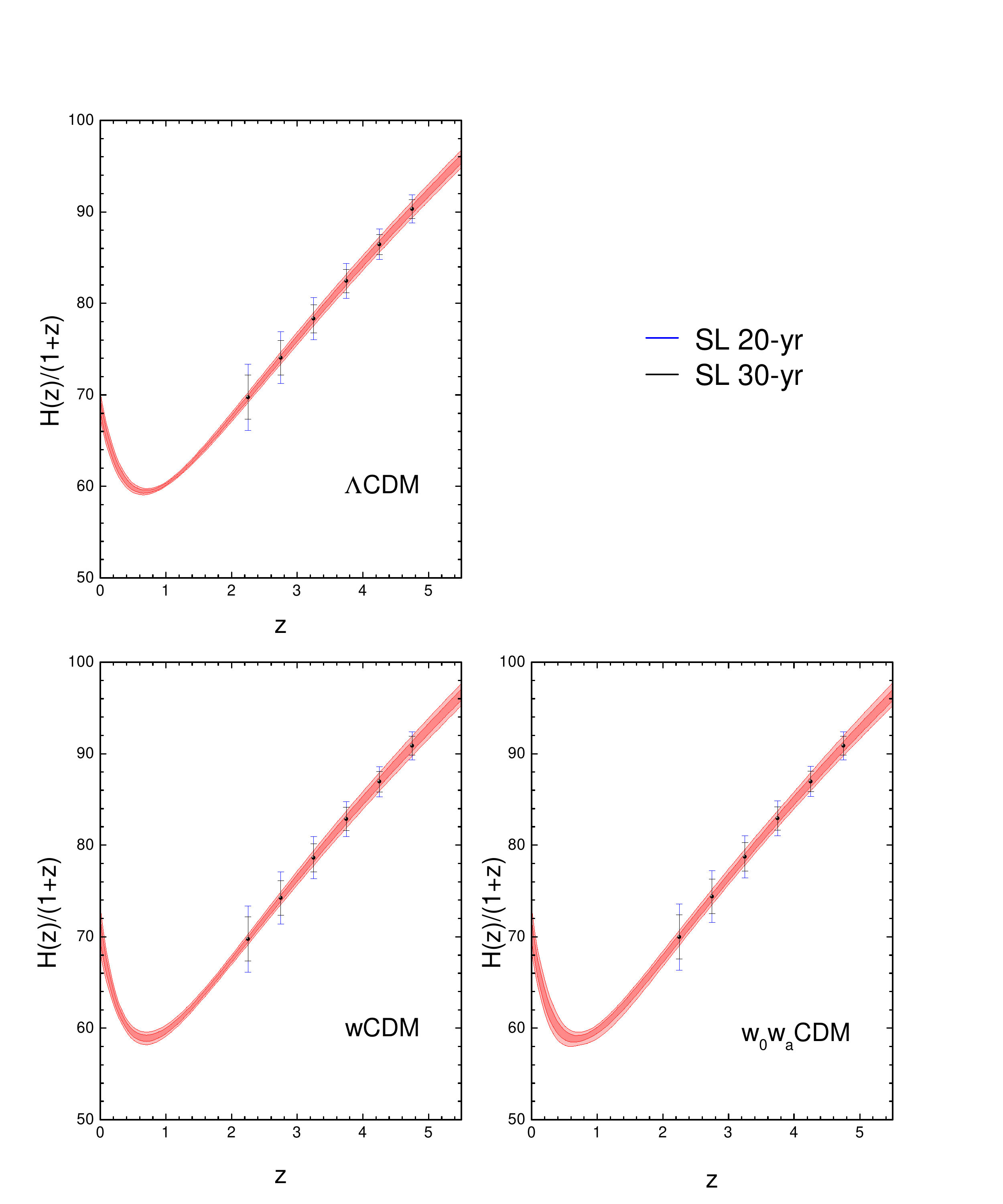}
\end{center}
\caption{Comparison of the $H(z)$ direct measurement data in SL test with the reconstructed $H(z)$ from the current SN+BAO+CMB+$H_0$ fit results.  1$\sigma$ and 2$\sigma$ uncertainties are shown.
Here $H(z)/(1+z)$ (instead of $H(z)$) is shown for a clearer display. }
\label{fig2}
\end{figure}

The SL test directly measures the redshift drifts in the range of $2\lesssim z\lesssim 5$; in other words, the SL test directly measures the Hubble expansion rate $H(z)$
at the high redshifts.
It is well known that the Hubble parameter $H(z)$ is related to the equation of state of dark energy through one integral [see Eq. (\ref{Ez}) for the integral $\int_0^z {1+w(z')\over 1+z'}dz'$ in $X(z)$], and the luminosity distance $d_L(z)$ (or the
angular diameter distance $d_A(z)$) is related to the equation of state of dark energy through two integrals
[both $d_L(z)$ and $d_A(z)$ are proportional to $D(z)=\int_0^z dz'/H(z')$ in a flat universe].

Thus, the direct measurements of $H(z)$ are of extreme
importance for constraining the property of dark energy.
If these high-redshift $H(z)$ data can be combined with some accurate low-redshift $H(z)$ data provided by other astrophysical methods, the capability of constraining dark energy
would be enormous.
Even though there are no such accurate low-redshift $H(z)$ data (there are indeed some low-redshift $H(z)$ data, but they are not sufficiently accurate \cite{Darling,Hz1,Hz2}), the high-redshift data
given by the SL test in combination with other low-redshift observations (such as SN and BAO) will also play a very significant role in constraining dark energy models.
Now, we wish to have a look at what accuracies the $H(z)$ measurements provided by SL test could achieve.
Hence, we plot the $H(z)$ evolutions for the three dark energy models in Fig.~\ref{fig2}.
In order to show the results more clearly, we actually plot the evolutions of $H(z)/(1+z)$ in this figure.
The red bands stand for the reconstructed $H(z)/(1+z)$ evolutions (with 1 and 2$\sigma$ uncertainties) for the three dark energy models from the fits to the current
SN+BAO+CMB+$H_0$ data, and the blue and black bars on the bands stand for the error bars of $H(z)/(1+z)$ measurements by 20-yr and 30-yr
observations of the SL test, respectively.
It can be seen from this figure that the accuracies of the SL high-redshift $H(z)$ data are not high compared to that of current data. However, in our following discussion, we will show that these SL high-redshift $H(z)$ data are extremely useful in breaking the significant degeneracies among the cosmological parameters in current data.

\begin{figure}
\begin{center}
\includegraphics[width=15cm]{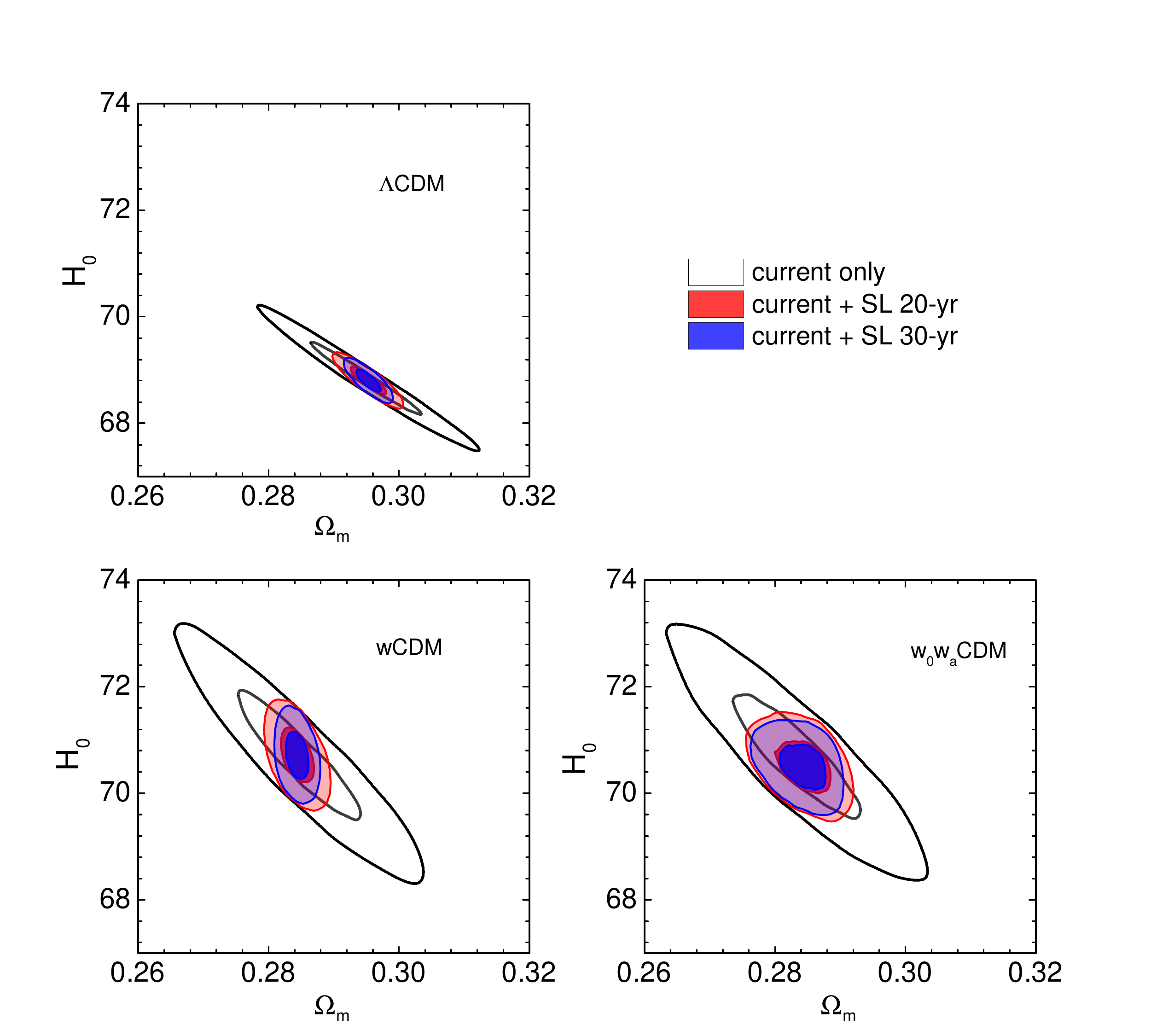}
\end{center}
\caption{Constraints (68.3\% and 95.4\% CL) in the $\Omega_m$--$H_0$ plane for $\Lambda$CDM, $w$CDM, and $w_0w_a$CDM models with current only, current+SL 20-yr and current+SL 30-yr data.}
\label{fig3}
\end{figure}

\begin{table*}\tiny
\caption{Errors of parameters in the $\Lambda$CDM ($\Lambda$), the $w$CDM ($w$), and the $w_0w_a$CDM ($w_0w_a$) models for the fits to
current only, current+SL 20-yr and current+SL 30-yr data.}
\label{table2}
\small
\setlength\tabcolsep{2.8pt}
\renewcommand{\arraystretch}{1.5}
\centering
\begin{tabular}{cccccccccccc}
\\
\hline\hline &\multicolumn{3}{c}{current only} &&\multicolumn{3}{c}{current + SL 20-yr} &&\multicolumn{3}{c}{current + SL 30-yr} \\
           \cline{2-4}\cline{6-8}\cline{10-12}
Error  & $\Lambda$ & $w$ & $w_0w_a$ & & $\Lambda$ & $w$ & $w_0w_a$& & $\Lambda$ & $w$ & $w_0w_a$\\ \hline

$\sigma(w_0)$              & $-$
                   & $0.082$
                   & $0.204$&
                   & $-$
                   & $0.068$
                   & $0.173$&
                   & $-$
                   & $0.062$
                   & $0.165$\\

$\sigma(w_a)$              & $-$
                   & $-$
                   & 0.767&
                   & $-$
                   & $-$
                   & $0.690$&
                   & $-$
                   & $-$
                   & $0.648$
                   \\

$\sigma(\Omega_{m})$       & $0.0125$
                   & $0.0140$
                   & $0.0146$&
                   & $0.0040$
                   & $0.0037$
                   & $0.0064$&
                   & $0.0028$
                   & $0.0026$
                   & $0.0052$\\

$\sigma(H_0)$              & $1.00$
                   & $1.81$
                   & $1.73$&
                   & $0.38$
                   & $0.76$
                   & $0.76$&
                   & $0.30$
                   & $0.64$
                   & $0.65$\\

\hline
\end{tabular}
\end{table*}

In the existing data, in particular the Planck CMB data, the strong degeneracy between $\Omega_m$ and $H_0$ is well known.
We shall show that the SL test data can effectively break this degeneracy and thus help constrain the parameters $\Omega_m$ and $H_0$
to a high precision.
Figure~\ref{fig3} shows the joint constraints on the $\Lambda$CDM, the $w$CDM, and the $w_0w_a$CDM models in the $\Omega_m$--$H_0$ plane.
The 68.3\% and 95.4\% CL posterior distribution contours are shown. The data combinations used are the current only, the current+SL 20-yr,
and the current+SL 30-yr combinations, and their constraint results are shown with white, red, and blue contours, respectively.
One can clearly see that the degeneracy between $\Omega_m$ and $H_0$ is well broken with the SL test data for all the three dark energy models.
The 1$\sigma$ errors of the parameters $w_0$, $w_a$, $\Omega_m$, and $H_0$ for the three models for the above three data combinations are given in Table~\ref{table2}.
From this table, one can directly figure out how the SL test data help improve the constraints.
With the 20-yr SL observation, the constraints on $\Omega_m$ and $H_0$ will be improved, respectively, by
68.0\% and 62.0\% for the $\Lambda$CDM model, by 73.6\% and 58.0\% for the $w$CDM model, and by $56.2\%$ and 56.1\% for the $w_0w_a$CDM model.
With the 30-yr SL observation, the constraints on $\Omega_m$ and $H_0$ will be improved, respectively, by
77.6\% and 70.0\% for the $\Lambda$CDM model, by 81.4\% and 64.6\% for the $w$CDM model, and by $64.4\%$ and 62.4\% for the $w_0w_a$CDM model.
Therefore, we can see that with a 30-yr observation of the SL test the geometric constraints on dark energy would be improved enormously.
For all the three dark energy models, the constraints on $\Omega_m$ and $H_0$ would be improved, relative to the current joint observations, by more than 60\%, with the SL 30-yr observation.

\begin{figure}
\begin{center}
\includegraphics[width=15cm]{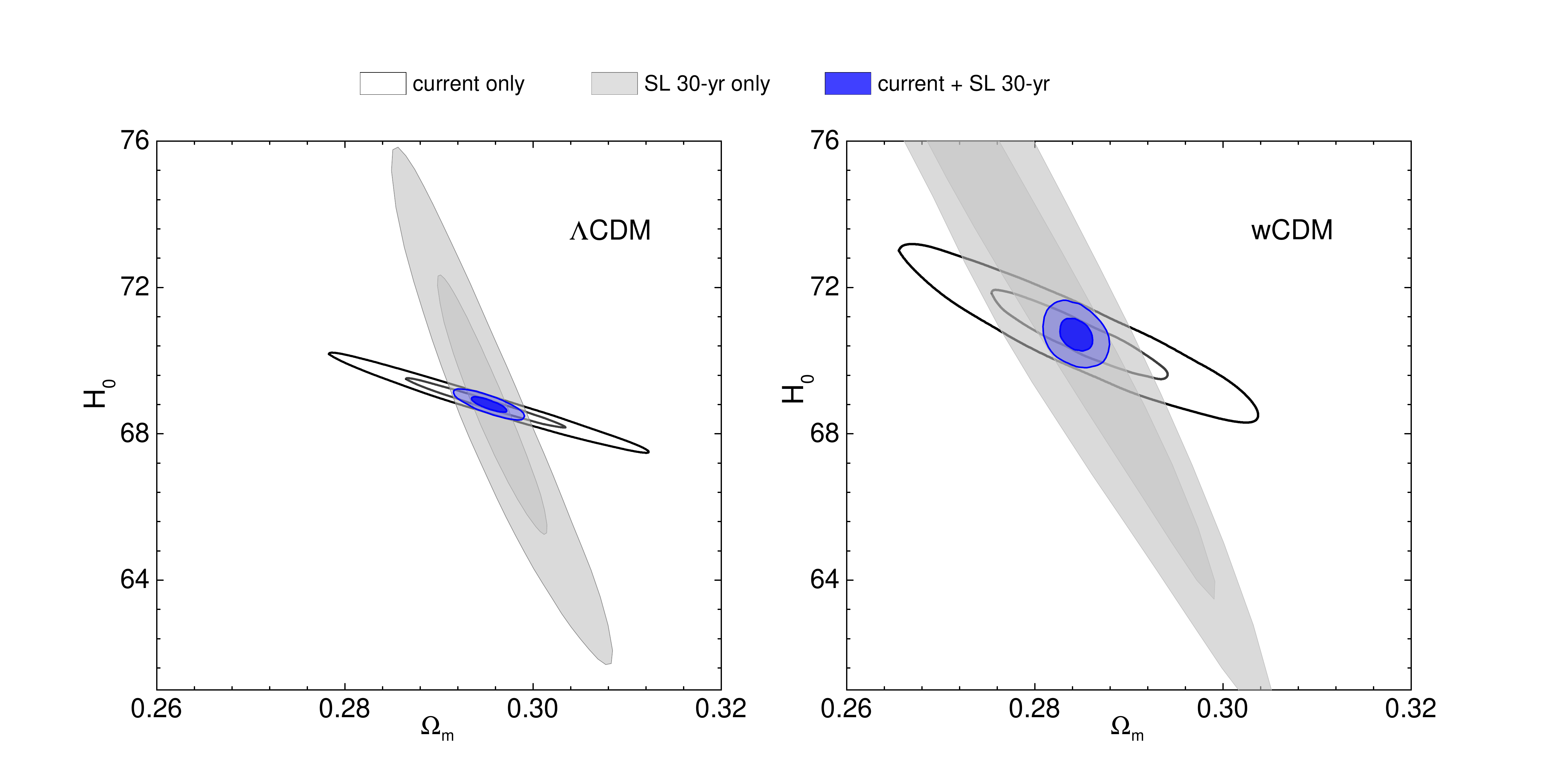}
\end{center}
\caption{ Constraints on the $\Lambda$CDM and the $w$CDM models in the $\Omega_m$--$H_0$ plane from the current only, SL 30-yr and current+SL 30-yr data. Owing to the fact that the orientations of the degeneracies in the two cases of current only and SL 30-yr constraints are very different, the strong degeneracy between  $\Omega_m$ and $H_0$ in the current data is well broken by the SL test.}
\label{fig4}
\end{figure}

Figure~\ref{fig4} shows how the SL test breaks the strong degeneracy between $\Omega_m$ and $H_0$ in the current data constraint.
Here we take the $\Lambda$CDM model and the $w$CDM model as examples.
The white contours are for the constraints from the current combined geometric observations, and the gray contours are for the constraints from the SL 30-yr only observation.
One can see clearly that the strong degeneracy between $\Omega_m$ and $H_0$ appears in both cases of the current only constraint and the SL 30-yr only constraint,
but the degeneracy orientations in the two cases are very different, and the strong degeneracy in the current data is thus well broken by the SL test.
The blue contours are for the results of the joint current+SL 30-yr data constraints, from which one can easily see that once the high-redshift SL test data are combined with the
current geometric observations the capability of constraining dark energy would be enhanced enormously.

\begin{figure}
\begin{center}
\includegraphics[width=15cm]{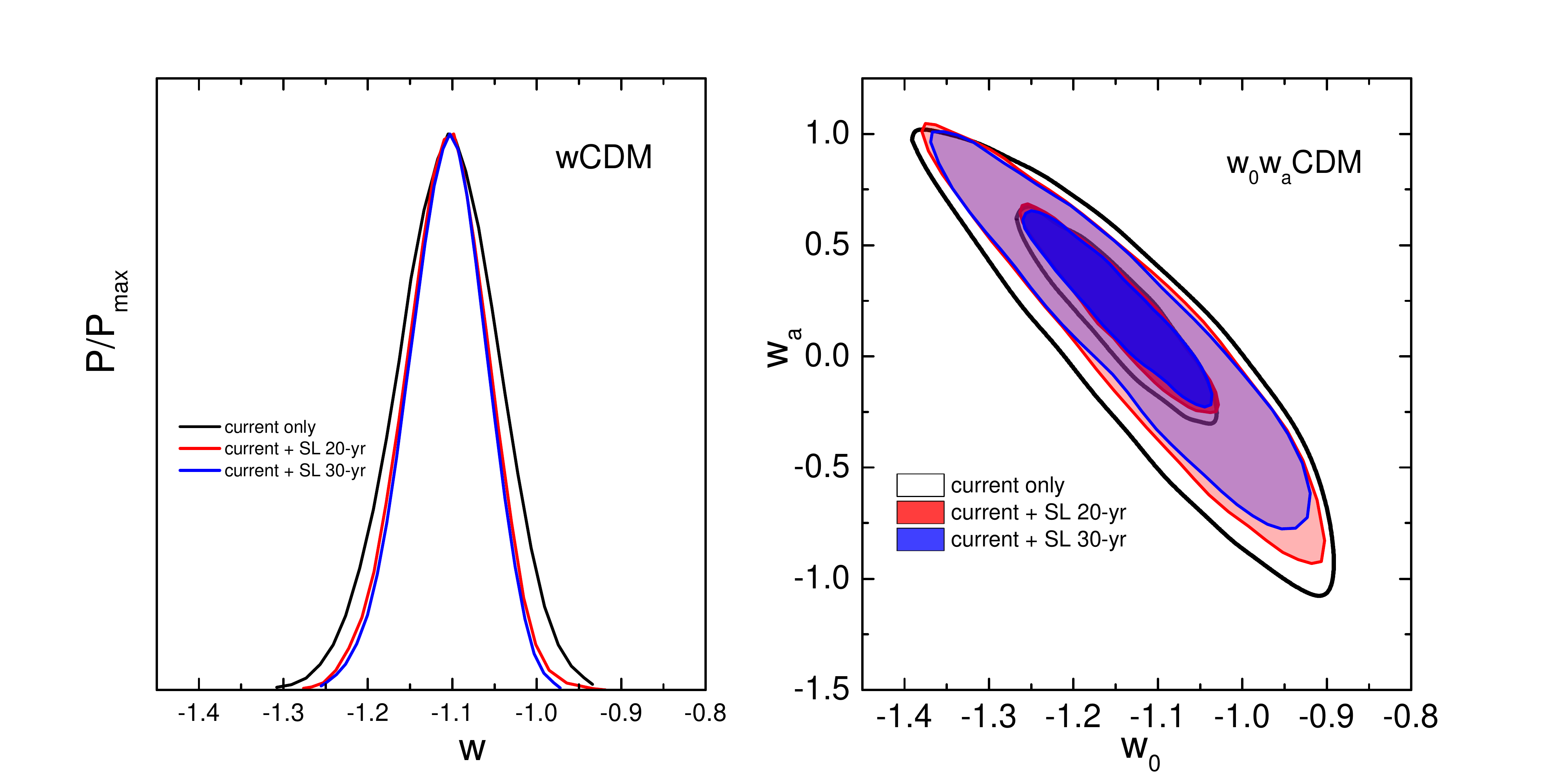}
\end{center}
\caption{The one-dimensional posterior distributions of $w$ for the $w$CDM model (left) and the two-dimensional posterior distributions of $w_0$ and $w_a$ for the $w_0w_a$CDM model (right), from the current only, current+SL 20-yr, and current+SL 30-yr constraints.}
\label{fig5}
\end{figure}

\begin{figure}
\begin{center}
\includegraphics[width=11cm]{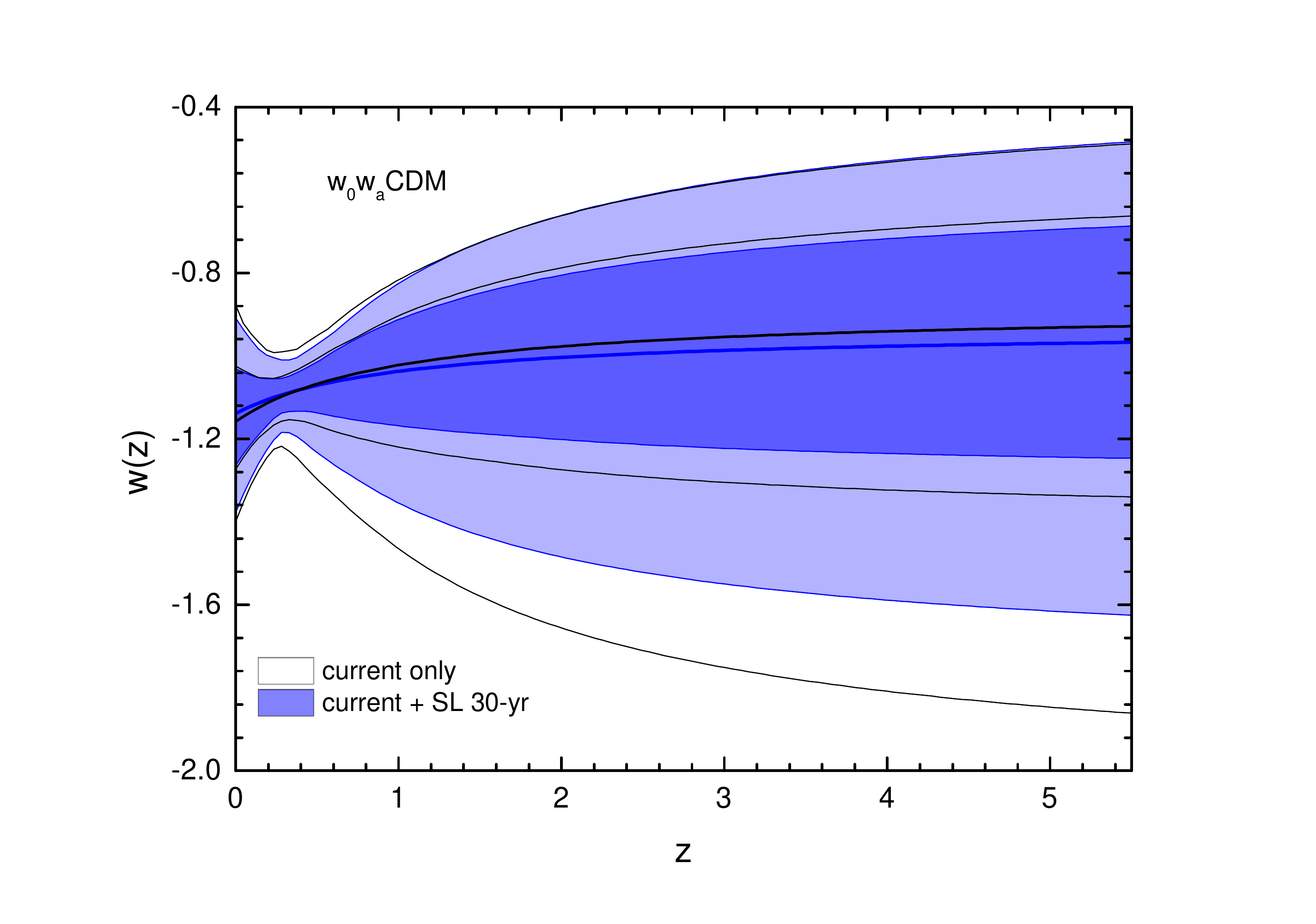}
\end{center}
\caption{Reconstructed $w(z)$ evolutions in the $w_0w_a$CDM model by using the constraint results of the current only and the current+SL 30-yr data. In this plot, 1$\sigma$ and 2$\sigma$ uncertainties are shown. }
\label{fig6}
\end{figure}

We also discuss the impact of the SL test data on constraining the dark energy equation of state. The case for the $w$CDM model has been discussed in Ref.~\cite{msl1}. In this paper,
we will analyze the case for the $w_0w_a$CDM model, and make a comparison for the two cases.
In Fig.~\ref{fig5} we show the one-dimensional posterior distributions of $w$ for the $w$CDM model and the two-dimensional posterior distributions of $w_0$ and $w_a$ for the
$w_0w_a$CDM model, from the current only, current+SL 20-yr, and current+SL 30-yr constraints.
The corresponding errors of $w_0$ and $w_a$ are given in Table~\ref{table2}.
For the $w$CDM model, the constraints on $w$ can be improved by 17.1\%, and 24.4\%, with 20-yr, and 30-yr observations, respectively.
For the $w_0w_a$CDM model, the SL 20-yr observation helps improve the constraints on $w_0$ and $w_a$ by 15.2\% and 10.0\%, respectively;
the SL 30-yr observation helps improve the constraints on $w_0$ and $w_a$ by 19.1\% and 15.5\%, respectively.
Therefore, we conclude that a 30-yr observation of the SL test can help improve the constraint on constant $w$ by about 25\%, and
improve the constraints on $w_0$ and $w_a$ by about 20\% and 15\%, respectively.
We also see that the SL test data cannot break the degeneracy between $w_0$ and $w_a$.
Furthermore, in Fig.~\ref{fig6} we reconstruct the $w(z)$ evolutions in the $w_0w_a$CDM model by using the constraint results of the current only and the current+SL 30-yr data.
From the comparison, we find that the SL test cannot greatly improve the reconstruction of $w(z)$.
In fact, the conclusion that the SL test could not break the current degeneracy between $w_0$ and $w_a$ has also been drawn in Ref. \cite{sl7}, and the possible reason has
been discussed in the same paper (for more details, see Sec. V of Ref. \cite{sl7}). However, we will see in the next section that the future high-precision SN and BAO observations could break the
degeneracy between $w_0$ and $w_a$ and measure both of them to a high precision, and in this case the SL test would further improve the measurement precision of $w_a$ by
more than 50\%.

\begin{figure}
\begin{center}
\includegraphics[width=13cm]{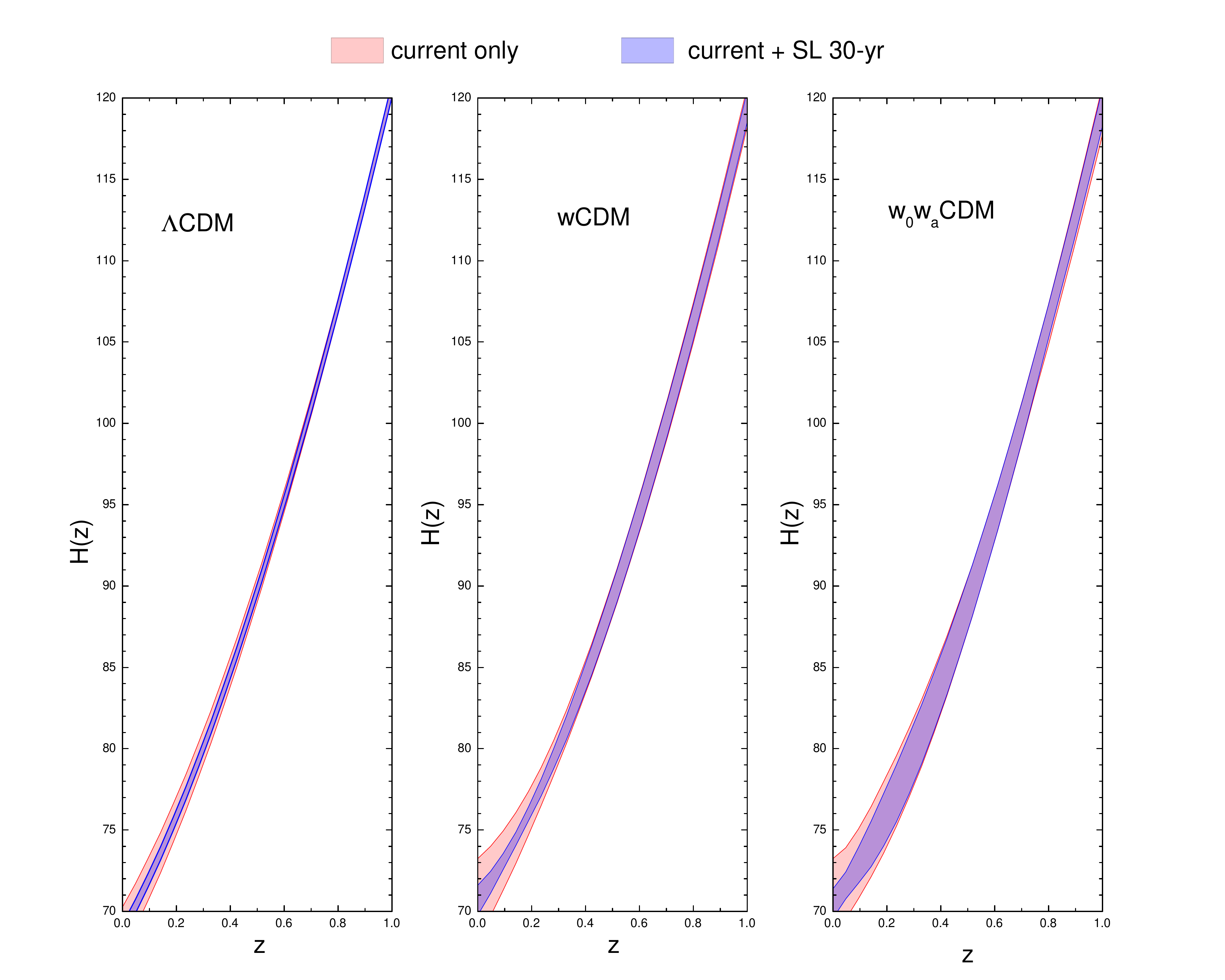}
\end{center}
\caption{Reconstructed $H(z)$ evolutions in the $\Lambda$CDM, the $w$CDM, and the $w_0w_a$CDM models, by using the
fit results from the current only and the current+SL 30-yr data. Only 1$\sigma$ uncertainties are shown in this plot.}
\label{fig7}
\end{figure}

We are also interested in the reconstruction of $H(z)$ with the SL test data.
In Fig.~\ref{fig7} we show the reconstructed $H(z)$ evolutions in the $\Lambda$CDM, the $w$CDM, and the $w_0w_a$CDM models, by using the
fit results from the current only and the current+SL 30-yr data.
From Fig.~\ref{fig2} we have learned that the accuracy of the high-redshift $H(z)$ direct measurements with the SL test is worse than that of the reconstructed $H(z)$ results from the
current combined data constraint. However, owing to the fact that the SL test data can break the parameter degeneracies in the current low-redshift geometric measurements,
the reconstructed $H(z)$ results are improved in the low redshifts with the help of the SL test data.
%

In the current constraints on dark energy, there are important parameter degeneracies, and we have shown that the future redshift-drift observations could play a crucial role in breaking these degeneracies. The next step is to test what role the redshift-drift measurements would play in the future combined geometric constraints. 

\section{Extended discussion concerning future geometric measurements}

In the above discussion, we showed how the future redshift-drift measurements would break the parameter degeneracies in the current geometric observations. However, when the CODEX experiment is ready to deliver its redshift-drift data in the future, other future geometric measurements data will also be available. Therefore, a further issue is to ask what role the SL test will play in improving the dark energy constraints in the future geometric measurements.

According to the report of the dark energy task force~\cite{DETF}, the most important future geometric measurements include the SN and the BAO observations (by the way, the most important structure growth measurements include the weak lensing and the galaxy clusters observations~\cite{DETF}). So in what follows we will only discuss the future long-term SN and BAO observations.

We simulate the future geometric measurements data using the method described in Ref.~\cite{DETF}.
As a concrete example, we simulate the future data based on the long-term space-based project called JDEM (Joint Dark Energy Mission) and simply describe the method in the following. For the details, we refer the reader to Ref.~\cite{DETF}.

For future SN data, aside from 2000 SNe distributed in 16 bins over the range from $z=0.1$ to $z=1.7$, a near sample of 500 SNe within $0.03<z<0.08$ is also considered.
The observables for SN data are apparent magnitudes $m_i=M+\mu(z_i)$, where $M$ represents the absolute magnitude, and $\mu(z_i)$ is the distance modulus given by 
$\mu(z_i)= 5 \log_{10} d_L(z_i)+25$.
The luminosity distance $d_L(z)=(1+z)\int_0^z dz'/H(z')$ for a flat universe.
The uncertainty of apparent magnitude $m_i$ due solely to variation in the properties of SN is denoted as $\sigma_D$. Besides, we use two additional nuisance parameters $\mu^L$ and $\mu^Q$ to give a quadratic $z$-dependent effect of the peak luminosity of SNe: $\mu(z_i)\rightarrow\mu(z_i)+\mu^Lz_i+\mu^Qz_i^2$. For the near sample, an additional nuisance parameter $\mu^S$ is included to represent an offset between the photometric systems of the near and far samples: $\mu(z_i)\rightarrow\mu(z_i)+\mu^S$. The uncertainties of these parameters used to simulate mock future data are $\sigma_D=0.10$, $\sigma_{\mu^L}=\sigma_{\mu^D}=0.01/\sqrt2$, and $\sigma_{\mu^S}=0.01$.

We simulate 10000 mock BAO data uniformly distributed among 10 redshift bins of $z\in [0.5,~2]$, with each $\Delta z_i$ centered on the grid $z_i$. The observables are expansion rate $H(z)$ and comoving angular diameter distance $d_A^{co}(z)=d_L(z)/(1+z)$.  The uncertainty of $\ln H(z_i)$ and $\ln d_A^{co}(z_i)$ can be expressed as
\begin{equation}
\sigma_H^i=x_0^H\frac{4}{3}\sqrt{\frac{V_0}{V_i}}f_{nl}(z_i),
\end{equation}
\begin{equation}
\sigma_d^i=x_0^d\frac{4}{3}\sqrt{\frac{V_0}{V_i}}f_{nl}(z_i),
\end{equation}
where the comoving survey volume in the redshift bin of $z_i$ is $V_i=1500(d_A^{co}(z_i))^2/H(z_i)$, and the erasure of the baryon features by non-linear evolution is factored in using $f_{nl}(z_i)=1$ for $z_i>1.4$ and $f_{nl}(z_i)=(1.4/z_i)^{1/2}$ for $z_i<1.4$. The parameters used in our simulation are $x_0^H=0.0148$, $x_0^d=0.0085$ and $V_0=\frac{2.16}{h^3}$ Gpc$^3$.
We also consider systematic errors in the BAO observation, which are modeled for both types of observable as independent uncertainties in the log of the distance measures in each redshift bin: $\sigma_s^i=0.01\sqrt{\frac{0.5}{\Delta{z_i}}}$, with $\Delta{z_i}=0.15$.

\begin{figure}
\begin{center}
\includegraphics[width=15cm]{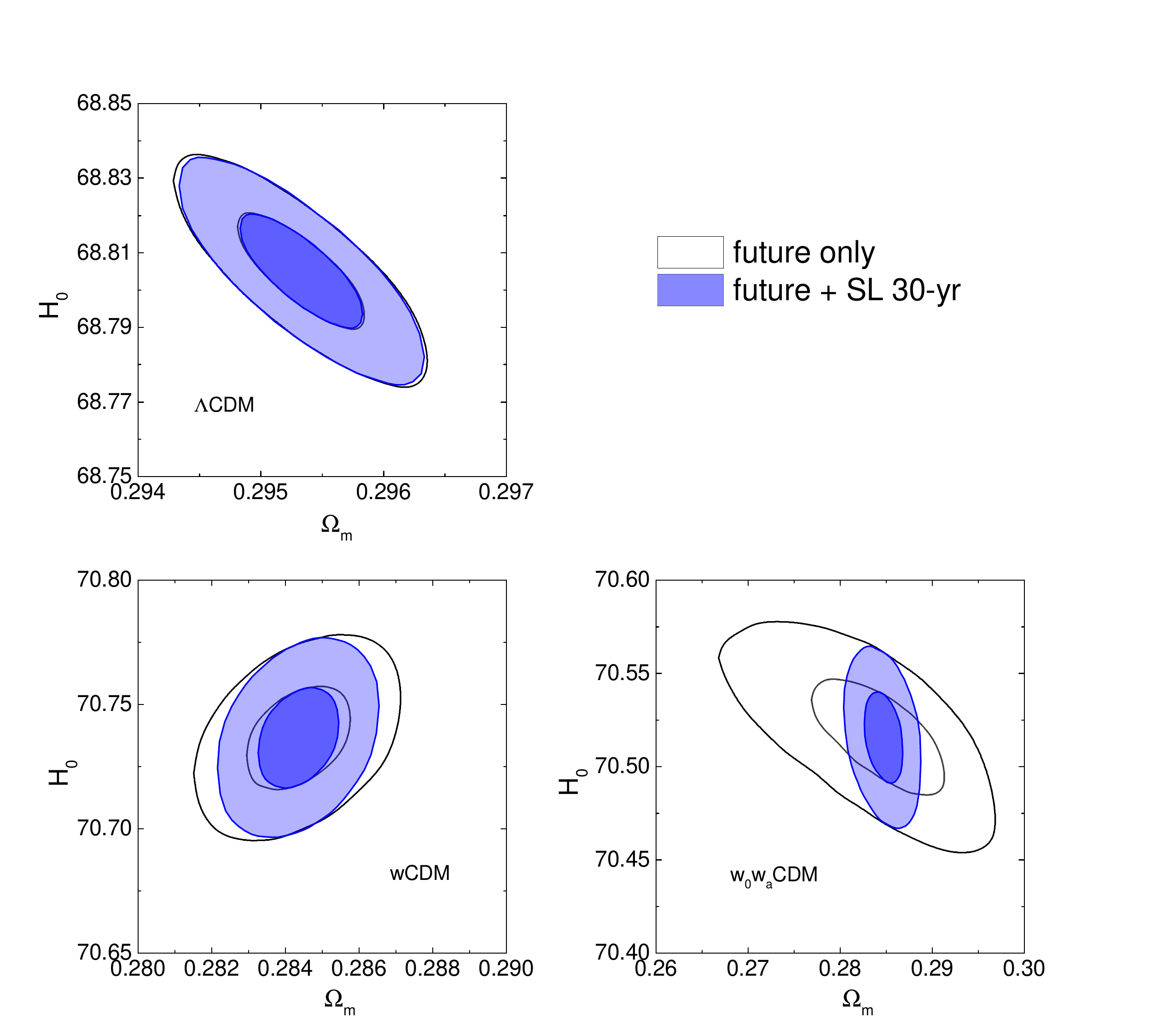}
\end{center}
\caption{Constraints (68.3\% and 95.4\% CL) in the $\Omega_m$--$H_0$ plane for $\Lambda$CDM, $w$CDM, and $w_0w_a$CDM models with future only and future + SL 30-yr data.}
\label{fig8}
\end{figure}

\begin{figure}
\begin{center}
\includegraphics[width=15cm]{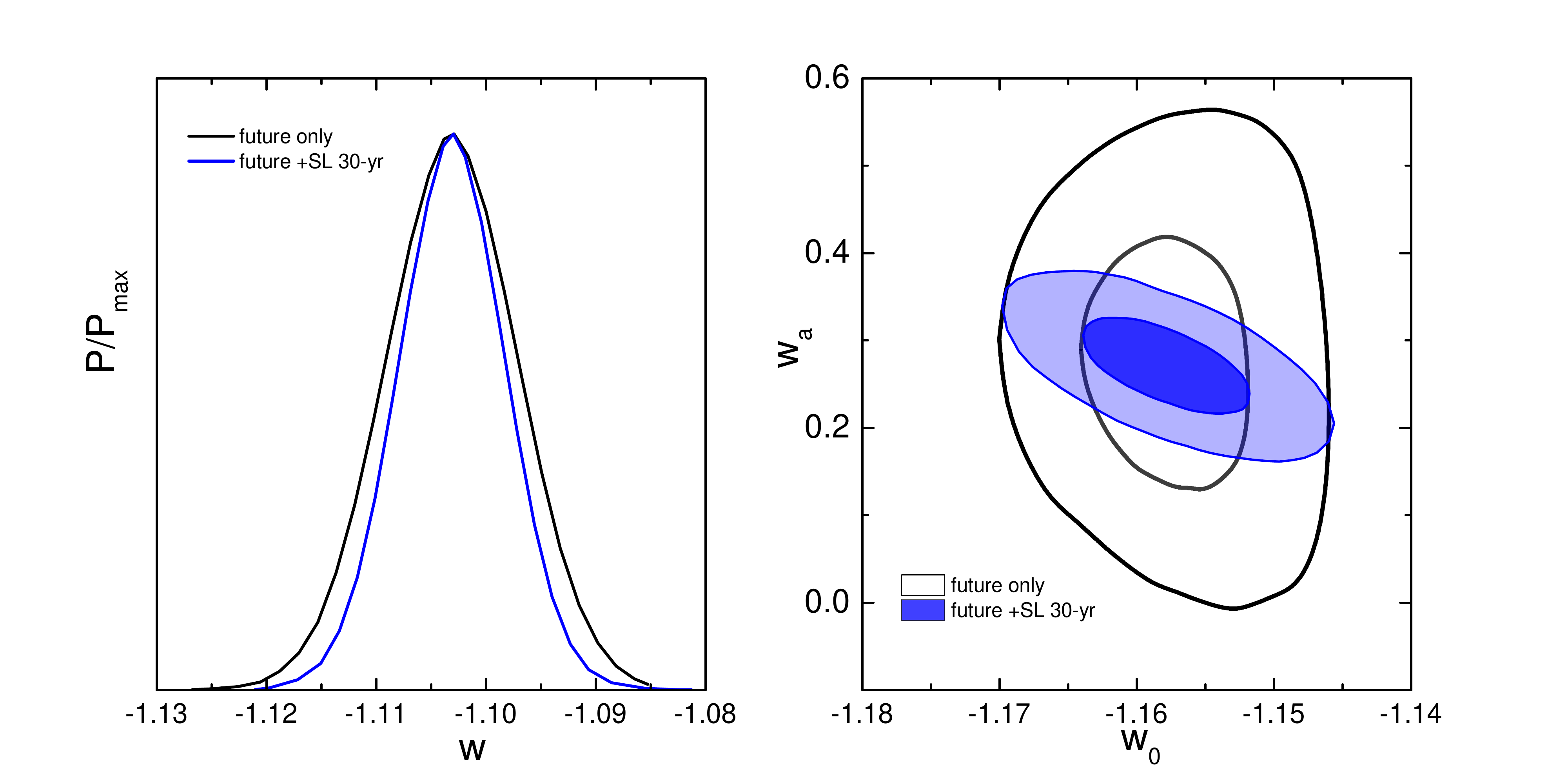}
\end{center}
\caption{The one-dimensional posterior distributions of $w$ for the $w$CDM model (left) and the two-dimensional posterior distributions of $w_0$ and $w_a$ for the $w_0w_a$CDM model (right), from the future only and the future + SL 30-yr constraints.}
\label{fig9}
\end{figure}

\begin{table*}\tiny
\caption{Errors of parameters in the $\Lambda$CDM ($\Lambda$), the $w$CDM ($w$), and the $w_0w_a$CDM ($w_0w_a$) models for the fits to
future only and future + SL 30-yr data.}
\label{table3}
\small
\renewcommand{\arraystretch}{1.5}
\centering
\begin{tabular}{ccccccccc}
\\
\hline\hline &\multicolumn{3}{c}{future only} &&\multicolumn{3}{c}{future + SL 30-yr} \\
           \cline{2-4}\cline{6-8}
Error  & $\Lambda$ & $w$ & $w_0w_a$ & & $\Lambda$ & $w$ & $w_0w_a$\\ \hline

$\sigma(w_0)$              & $-$
                   & $0.0083$
                   & $0.0091$&
                   & $-$
                   & $0.0067$
                   & $0.0090$\\

$\sigma(w_a)$              & $-$
                   & $-$
                   & $0.208$&
                   & $-$
                   & $-$
                   & $0.079$
                   \\

$\sigma(\Omega_{m})$       & $0.00078$
                   & $0.0021$
                   & $0.0108$&
                   & $0.00075$
                   & $0.0016$
                   & $0.0031$\\

$\sigma(H_0)$              & $0.0223$
                   & $0.0296$
                   & $0.0449$&
                   & $0.0218$
                   & $0.0286$
                   & $0.0349$\\
\hline
\end{tabular}
\end{table*}

Figure~\ref{fig8} shows the joint constraints on the $\Lambda$CDM, the $w$CDM, and the $w_0w_a$CDM models in the $\Omega_m$--$H_0$ plane.
The 68.3\% and 95.4\% CL posterior distribution contours are shown. The data combinations used are the future only and the future + SL 30-yr combinations, and their constraint results are shown with white and blue contours, respectively.
The 1$\sigma$ errors of the parameters $w_0$, $w_a$, $\Omega_m$, and $H_0$ for the three models for the above data combinations are given in Table~\ref{table3}.
Note that here we use ``future'' to denote the data combination of future SN and BAO for convenience.
It is shown that with the 30-yr SL observation, the constraints on $\Omega_m$ and $H_0$ will be improved by
3.8\% and 2.2\% for the $\Lambda$CDM model, by 23.8\% and 3.4\% for the $w$CDM model, and by $71.3\%$ and 22.3\% for the $w_0w_a$CDM model.

In Fig.~\ref{fig9}, we show the one-dimensional posterior distributions of $w$ for the $w$CDM model (left panel) and the two-dimensional posterior distributions of $w_0$ and $w_a$ for the
$w_0w_a$CDM model (right panel), from the future only and the future + SL 30-yr constraints.
The corresponding 1$\sigma$ errors of $w_0$ and $w_a$ are given in Table~\ref{table3}.
For the $w$CDM model, the constraints on $w$ can be improved by 19.3\%, with the SL 30-yr observation.
For the $w_0w_a$CDM model, the SL 30-yr observation helps improve the constraints on $w_0$ and $w_a$ by 1.1\% and 62.0\%, respectively.

Therefore, we find that the redshift-drift observation could also play an important role in improving the dark energy constraints from the future geometric measurements,
especially for the $w_0w_a$CDM model. In the future geometric constraints for the $w_0w_a$CDM model, the SL 30-yr observation would help improve the measurement precisions of
$\Omega_m$, $H_0$, and $w_a$ by more than 70\%, 20\%, and 60\%, respectively.

\section{Summary}

In this paper, we have discussed the parameter estimation for the $\Lambda$CDM, the $w$CDM, and the $w_0w_a$CDM models with the future SL test data.
The SL test directly measures the expansion rate of the universe in the redshift range of $2\lesssim z \lesssim 5$ by detecting redshift drift in the spectra of Lyman-$\alpha$ forest
of distant QSOs, thus as a purely geometric measurement it provides an important supplement to other dark energy probes.
Following our previous work \cite{msl1}, in order to guarantee that the simulated SL test data are consistent with the other geometric measurement data,
we used the best-fitting dark energy models constrained by the current combined geometric measurement data as the fiducial models to produce the mock SL test data and
then used these simulated data to do the analyses.

We showed that the SL test data are extremely helpful in breaking the existing parameter degeneracies.
The strong degeneracy between $\Omega_m$ and $H_0$ in the current SN + BAO + CMB + $H_0$ constraint results for all the three models can be well broken by the SL test.
By analyzing and comparing the 20-yr and 30-yr observations of SL test, we found that the 30-yr observation could provide sufficiently important supplement to the other
observations. Compared to the current SN + BAO + CMB + $H_0$ constraint results, the 30-yr observation of SL test could improve the constraints on
$\Omega_m$ and $H_0$ by more than 60\% for all the three models.
But the SL test can only moderately improve the constraint on the equation of state of dark energy.
We showed that a 30-yr observation of SL test could help improve the constraint on constant $w$ by about 25\%, and improve the constraints on $w_0$ and $w_a$ by
about 20\% and 15\%, respectively.

We also analyzed how the SL test would impact on the dark energy constraints from the future geometric measurements. To do this analysis, we simulated the future SN and BAO
data based on the long-term space-based project JDEM. We found that the SL test could play a crucial role in the future joint geometric constraints. For example, the 30-yr observation of
SL test would help improve the measurement precision of $\Omega_m$, $H_0$, and $w_a$ by more than 70\%, 20\%, and 60\%, respectively, for the $w_0w_a$CDM model.

As a purely geometric measurement, the SL test has been proven to be a very important supplement to the other geometric measurement observations.
Actually, in order to differentiate the noninteracting dark energy, interacting dark energy, and modified gravity models, the geometric measurements should be compared to
the measurements of the growth of large-scale structure. A consistency test of the geometric and structural measurements might provide a diagnostic to the cause of the
acceleration of the universe in the future. Of course, the SL test will definitely play a significant role in doing such an analysis.
For the interacting dark energy models, the longstanding problem of large-scale instability was recently resolved by establishing a parameterized post-Friedmann framework
for interacting dark energy~\cite{PPF,PPF2}. Thus, the interacting dark energy models with the background interaction forms of both $Q\propto \rho_c$ and $Q\propto \rho_{de}$ are now
proven to be well behaved. It is, undoubtedly, worthy to study the interacting dark energy models with the SL test. In Ref.~\cite{msl1}, a preliminary SL test analysis has been
made for the constant $w$ model with $Q=\gamma H\rho_c$ and $Q=\gamma H\rho_{de}$. However, an analysis for the models with $Q=\Gamma \rho_c$ and $Q=\Gamma \rho_{de}$
(here $\Gamma$ is a constant) is still absent. We will leave the complete analysis for interacting dark energy models and modified gravity models in future work.

\acknowledgments

We acknowledge the use of {\tt CosmoMC}.We thank Yun-He Li for helpful discussion.
JFZ is supported by the Provincial Department of Education of
Liaoning under Grant No.~L2012087.
XZ is supported by the National Natural Science Foundation of
China under Grant No.~11175042 and the Fundamental Research Funds for the
Central Universities under Grant No.~N120505003.




\end{document}